\documentclass[aps,twocolumn,superscriptaddress,showpacs]{revtex4}
\usepackage{epsfig}
\usepackage{times}

\begin{document}

\title{Culturomics meets random fractal theory: Insights into long-range correlations of social and natural phenomena over the past two centuries}

\author{Jianbo Gao}
\email{jbgao.pmb@gmail.com}
\affiliation{PMB Intelligence, LLC, West Lafayette, IN 47996, USA}
\affiliation{Department of Mechanical and Materials Engineering, Wright State University, Dayton, OH 45435, USA}

\author{Jing Hu}
\affiliation{Affymetrix, Inc., 3380 Central Expressway, Santa Clara, CA 95051, USA}

\author{Xiang Mao}
\affiliation{Department of Electrical and Computer Engineering, University of Florida, Gainesville, FL 32611, USA}

\author{Matja{\v z} Perc}
\email{matjaz.perc@uni-mb.si}
\affiliation{Faculty of Natural Sciences and Mathematics, University of Maribor,
Koro{\v s}ka cesta 160, SI-2000 Maribor, Slovenia}

\begin{abstract}
Culturomics was recently introduced as the application of high-throughput data collection and analysis to the study of human culture. Here we make use of this data by investigating fluctuations in yearly usage frequencies of specific words that describe social and natural phenomena, as derived from books that were published over the course of the past two centuries. We show that the determination of the Hurst parameter by means of fractal analysis provides fundamental insights into the nature of long-range correlations contained in the culturomic trajectories, and by doing so, offers new interpretations as to what might be the main driving forces behind the examined phenomena. Quite remarkably, we find that social and natural phenomena are governed by fundamentally different processes. While natural phenomena have properties that are typical for processes with persistent long-range correlations, social phenomena are better described as nonstationary, on-off intermittent, or L{\'e}vy walk processes.
\end{abstract}

\maketitle

\section*{1. INTRODUCTION}

Observational data are often very complex, appearing without any structure or pattern in either time or space. Examples of such observations can be found across the whole spectrum of the social and natural sciences, ranging from economics \cite{mantegna_00}, to physics \cite{kantz_98}, biology \cite{glass_88}, and medicine \cite{goldberger_c00}. The origins of observed irregular behavior, however, are not always clear. Roughly five decades ago, deterministic chaos was discovered \cite{lorenz_jas63} and quickly rose to prominence as a possible mechanism of inherent unpredictability and complexity \cite{crutchfield_pr82,eckmann_rmp85}. Yet the strict criteria for declaring deterministic chaos in observed data \cite{abarbanel_96}, most notably the satisfaction of criteria for stationarity and determinism \cite{kantz_98}, and the verification of exponential divergence \cite{gao_07,gao_pre06}, are rarely satisfied. In response, attention has begun to shift from chaos to noise and random processes as alternate \cite{stratonovich_63} (or, in many cases, as even more probable) sources of irregularity. While the theory of deterministic chaos relies on nonlinear dynamical systems with typically only a few degrees of freedom, the analysis of stochastic processes, especially those that yield data with scale invariance, relies on random fractal theory \cite{mandelbrot_82} or its generalization, multifractal theory \cite{bunde_96,gao_07}. Indeed, investigations based on these theoretical foundations may provide an elegant statistical characterization of a broad range of heterogeneous phenomena \cite{stanley_n88}, and in this paper, it is our goal to extend this theory to culturomics, as recently introduced in \cite{michel_s11}.

Culturomics, and the study of human culture in general, seemingly has little to do with deterministic chaos and fractals. However, quantitative analyses of various aspects of human culture have become increasingly popular; examples include the study of human mobility patterns \cite{gonzales_n08,song_np10,song_s10}, the spread of infectious diseases \cite{balcan_pnas09,meloni_pnas09,sanz_pre10,meloni_sr11} and malware \cite{hu_pnas09,wang_s09}, the dynamics of online popularity \cite{ratkiewicz_prl10}, social movement \cite{holt_pone11} and language \cite{lieberman_n07,puglisi_pnas08,loreto_jsm11}, and even tennis \cite{radicchi_po11}. This progress is driven not only by important advances in theory and modeling, but also by the increasing availability of vast amounts of data and knowledge, also referred to as metaknowledge \cite{evans_s11}, which allows scientists to apply advanced methods of analysis on a large scale \cite{lazer_s09}. The seminal study by \cite{michel_s11} was accompanied by the release of a vast amount of data comprised of metrics derived from $\sim4\%$ of books ever published (over five million in total), and it was this release that made the present study, \textit{i.e.} the application of random fractal theory, possible. The data are available at ngrams.googlelabs.com as counts of $n$-grams that appeared in a given corpus of books published in each year.  An $n$-gram is made up of a series of $n$ $1$-grams, and a $1$-gram is a string of characters uninterrupted by a space. Note that a $1$-gram is not necessarily a word, for it may be a number or a typo as well. Besides the counts of individual $n$-grams, the total counts of $n$-grams contained in each corpus of books in a given year are also provided, from which yearly usage frequencies can be obtained.

In this paper, we show what new insights are attainable by applying random fractal theory to this vast culturomic data set. Our goal is to try and go beyond the interpretations of trajectories provided in \cite{michel_s11} by means of an accurate determination of scaling parameters \cite{stanley_87}, and in particular the Hurst parameter $H$, which enables us to characterize the nature of correlations (memory), if any, contained in the irregular time series. In general, data with long-range correlations are an important subclass of $1/f^\alpha$ noise \cite{press_cmpc78,bak_prl87,bak_96}, which is characterized by a power-law decaying power spectral density, and whose dimensionality cannot be reduced by principal component analysis since the rank-ordered eigenvalue spectrum also decays as a power law \cite{gao_pla03}. Processes that generate time series with such properties are said to have antipersistent correlations if $0<H<1/2$, are memoryless or have only short-range correlations if $H=1/2$, and have persistent long-range correlations (long memory) if $1/2<H<1$ \cite{mandelbrot_82}. Moreover, values of $H>1$ are possible as well; these values, however, are characteristic of nonstationary processes or rather special stationary processes such as on-off intermittency with power-law distributed on and/or off periods and L{\'e}vy walks \cite{gao_pre06}. (Note that the latter should not be confused with L{\'e}vy flights, which are random processes consisting of many independent steps, and are thus memoryless with $H=1/2$.) Prominent examples where $1/f^\alpha$ noise was recently observed and quantified include DNA sequences \cite{voss_prl92,peng_n92}, human cognition \cite{gilden_s95} and coordination \cite{chen_prl97}, posture \cite{collins_prl94}, cardiac dynamics \cite{ivanov_n96,amaral_prl98,ivanov_n99,galvan_prl01}, as well as the distribution of prime numbers \cite{wolf_pa97}, to name but a few.

Despite the many successful attempts at assessing long-range correlations in complex time series -- for example, by means of detrended fluctuation analysis \cite{peng_pre94}, as well as many other methods \cite{bunde_96,gao_07} -- care should be exercised by their interpretation, particularly if one is faced with relatively short time series that contain trends \cite{hu_pre01}, nonstationarity \cite{stanley_pa02}, or signs of rhythmic activity \cite{chen_pre05,hu_jsm09}. Although it is obviously impossible to make general statements concerning these properties for all the $n$-grams contained in the corpus of the over five million digitized books, which amount roughly to over two billion culturomic trajectories, it is clear that the time series are short, comprising a little  more than $\sim200$ points corresponding to the two centuries considered (more precisely, from year 1770 to 2007), and that many will inevitably contain strong trends \cite{michel_s11}. In order to successfully surpass the difficulties and pitfalls associated with the analysis of such time series \cite{gao_pre06}, besides the traditional detrended fluctuation analysis (DFA), we also use an adaptive fractal analysis (AFA), which is based on nonlinear adaptive multiscale decomposition. We use these methods to determine the Hurst parameter $H$ for several $1$-grams that are representative for social and natural phenomena. Examples of words that we focus on include war, unemployment, hurricane and earthquake (see Tables~1 and 2 for the complete list), and we find that those that describe social phenomena (war, unemployment, etc.) in general have different scaling properties than those describing natural phenomena (hurricane, earthquake, etc.). Our results can be corroborated aptly with arguments from real life, and they fit nicely to the declared goal of culturomics, which is to extend the boundaries of scientific inquiry to a wide array of new phenomena \cite{michel_s11}.

The remainder of this paper is organized as follows. In the next section we present the main results, in section 3 we summarize them and discuss their potential implications, while in the appendix we describe the details of fractal analysis.

\section*{2. RESULTS}
\subsection*{2.1. Natural phenomena}

\begin{figure}
\begin{center}
\includegraphics[width=8.1cm]{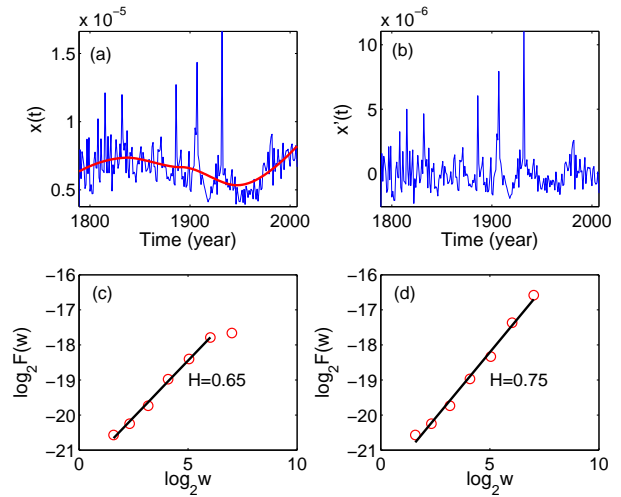}
\end{center}
\caption{Adaptive fractal analysis of the usage frequency of the 1-gram ``earthquake'' in the corpus of English books. The Hurst parameter, as obtained from the detrended data, is $H=0.65$. a) The variation of the usage frequency of ``earthquake'' with time. The blue (thin) line depicts original data, while the red (thick) line depicts the estimated trend (using a window of length $101$). b) Detrended data, \textit{i.e.,} the difference between the blue and red curves in panel a). c) Best fit to the $F(w)$ versus $w$ dependence for detrended data on a double log scale yields $H=0.65$. d) Best fit to the $F(w)$ versus $w$ dependence for original data on a double log scale yields $H=0.75$.}
\label{earthquake}
\end{figure}

We start by presenting the results of the adaptive fractal analysis for natural phenomena. In figure~\ref{earthquake}, we first plot in panel (a) the original time series (thin blue line) and the estimated trend (thick red line) for the 1-gram ``earthquake''. The detrended data are presented in panel (b). It can be observed that overall the trend is very modest and simple, increasing only slightly towards the present day. Using equation~\ref{ada}, the Hurst parameter can be estimated from the slope of the $F(w)$ versus $w$ dependence on a double log scale. In panel (c), we show that the analysis of detrended data yields $H=0.65$, while in panel (d), we show that $H=0.75$ if the original data is used as input. Both calculations produce similar results, showing a very modest slope, and rely on statistically robust scaling. Based on the meaning of the Hurst parameter, the fractal analysis of the culturomic trajectory for ``earthquake'' reveals that this phenomenon has persistent long-range correlation.

\begin{figure}
\begin{center}
\includegraphics[width=8.1cm]{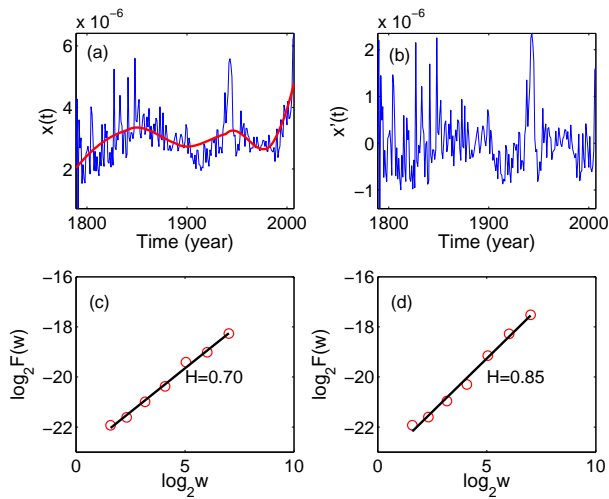}
\end{center}
\caption{Adaptive fractal analysis of the usage frequency of the 1-gram ``hurricane'' in the corpus of English books. The Hurst parameter, as obtained from the detrended data, is $H=0.70$. a) The variation of the usage frequency of ``hurricane'' with time. The blue (thin) line depicts the original data, while the red (thick) line depicts the estimated trend (using a window of length $101$). b) Detrended data, \textit{i.e.} the difference between the blue and red curves in panel a). c) Best fit to the $F(w)$ versus $w$ dependence for detrended data on a double log scale yields $H=0.70$. d) Best fit to the $F(w)$ versus $w$ dependence for original data on a double log scale yields $H=0.85$.}
\label{hurricane}
\end{figure}

As another example, we show in figure~\ref{hurricane} the same analysis for the 1-gram ``hurricane''. Unlike the ``earthquake'' trajectory, the trend for ``hurricane'' is more pronounced. It has a strong upwards component, especially in the last couple of decades. Hence, it can be expected that the discrepancy of the two estimated $H$ values for the original and detrended data will be somewhat larger than that for the 1-gram ``earthquake'' analyzed in figure~\ref{earthquake}. This expectation is indeed confirmed by comparing panels (c) and (d), from where it follows that for the detrended data $H=0.70$ while for original data $H=0.85$. Still, however, both results robustly classify ``hurricane'' as a phenomenon with persistent long-range correlations, thus adding to the evidence that this may be valid in general for natural phenomena.

To test this hypothesis more thoroughly, we have performed the same analysis as depicted in figures~\ref{earthquake} and \ref{hurricane}, along with the detrended fluctuation analysis (DFA), for thirteen other phenomena that can be classified as characteristic of natural phenomena. Although there may be some disagreement as to what terms are \textit{characteristic} of natural phenomenon, and other 1-grams as well as $n$-grams could be suggested as characteristic of natural phenomena and analyzed, we consider our selection to be sufficiently representative for this study. Supporting this assumption are the results presented in table~\ref{natural}, which point robustly towards the conclusion that natural phenomena in general really can be classified as processes with persistent long-range correlations. More specifically, for detrended data, we find that all estimated Hurst parameters are within the $1/2<H<1$ range with an average of $\overline H=0.69$ (AFA), which leads us to the mentioned final conclusion. Results obtained for original data (before detrending, not shown), on the other hand, leave a bit more room for discussion. There, for certain 1-grams, like ``mudslide'' and ``flooding'', the value of $H$ is larger than one. This suggests that the data would be more appropriately described as being either nonstationary, on-off intermittent, or L{\'e}vy walk-like. Such a discussion, however, would be to a large degree baseless as the upward trends occurring towards the present time in most $n$-grams describing natural phenomena must be properly taken into account. The observed trends may be considered as a straightforward consequence of the fact that we have more and more data readily available on natural phenomena, which is due to advancements in measuring techniques as well as the increasingly global reach of the Internet. Modern data collection and telecommunication technologies have raised our awareness, in general, of natural phenomena, and as a result, it is reasonable to expect this increased awareness to be reflected in an increase of occurrences in the corpus. Note, however, that similar arguments can be raised for other fields and trivia (\textit{e.g.} celebrity gossip, popular culture) as well, and thus one could argue that relatively, the usage frequencies should not necessarily increase as a result of that.

\begin{table}
\begin{center}
\begin{tabular}{|c|c|c|}
\multicolumn{3}{c}{}\\\hline
1-grams & \multicolumn{2}{c|} {Hurst Parameter ($H$)}  \\\cline{2-2}\cline{3-3}
                & AFA             & DFA             \\\hline
avalanche       & 0.63 $\pm$ 0.06 & 0.79 $\pm$ 0.06 \\\hline
comet           & 0.60 $\pm$ 0.03 & 0.73 $\pm$ 0.04 \\\hline
drought         & 0.81 $\pm$ 0.05 & 0.69 $\pm$ 0.09 \\\hline
earthquake      & 0.65 $\pm$ 0.02 & 0.72 $\pm$ 0.03 \\\hline
erosion         & 0.85 $\pm$ 0.06 & 0.86 $\pm$ 0.08 \\\hline
fire            & 0.67 $\pm$ 0.05 & 0.70 $\pm$ 0.03 \\\hline
flooding        & 0.85 $\pm$ 0.06 & 0.72 $\pm$ 0.08 \\\hline
hurricane       & 0.70 $\pm$ 0.03 & 0.69 $\pm$ 0.08 \\\hline
landslide       & 0.66 $\pm$ 0.05 & 0.41 $\pm$ 0.20 \\\hline
life            & 0.62 $\pm$ 0.03 & 0.65 $\pm$ 0.06 \\\hline
lightning       & 0.63 $\pm$ 0.03 & 0.70 $\pm$ 0.03 \\\hline
mudslide        & 0.80 $\pm$ 0.02 & 0.58 $\pm$ 0.28 \\\hline
tornado         & 0.59 $\pm$ 0.02 & 0.64 $\pm$ 0.06 \\\hline
tsunami         & 0.81 $\pm$ 0.05 & 0.66 $\pm$ 0.03 \\\hline
typhoon         & 0.55 $\pm$ 0.02 & 0.50 $\pm$ 0.09 \\\hline
\end{tabular}
\end{center}
\caption{\label{natural}Hurst parameters $H$, as obtained for the detrended data of all fifteen considered 1-grams describing natural phenomena. The left column lists results as obtained with the adaptive fractal analysis (AFA), while the right column lists results as obtained with the detrended fluctuation analysis (DFA). The range of values as obtained by AFA is $0.55 \leq H \leq 0.85$, with an average over all fifteen considered 1-grams equalling $\overline H=0.69$. With DFA we obtain $0.41 \leq H \leq 0.85$ and $\overline H=0.67$.}
\end{table}

\subsection*{2.2. Social phenomena}

\begin{figure}
\begin{center}
\includegraphics[width=8.1cm]{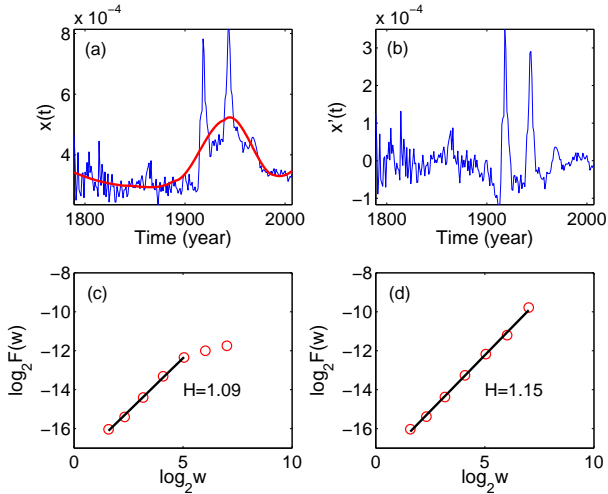}
\end{center}
\caption{Adaptive fractal analysis of the usage frequency of the 1-gram ``war'' in the corpus of English books. The Hurst parameter, as obtained from the detrended data, is $H=1.09$. a) The variation of the usage frequency of ``war'' with time. The blue (thin) line depicts the original data, while the red (thick) line depicts the estimated trend (using a window of length $101$). b) The detrended data, \textit{i.e.} the difference between the blue and red curves in panel a). c) Best fit to the $F(w)$ versus $w$ dependence for detrended data on a double log scale yields $H=1.09$. d) Best fit to the $F(w)$ versus $w$ dependence for original data on a double log scale yields $H=1.15$.}
\label{war}
\end{figure}

Turning to social phenomena, we will show that the problems discussed for natural phenomena are in some cases amplified, but more importantly, that social phenomena, apart from rare exceptions, cannot be classified solely as processes with persistent-long range correlations.

First, we presented the adaptive fractal analysis for the 1-gram ``war'' in figure~\ref{war}. The original data depicted by the thin blue line in panel (a) are clearly reminiscent of historical events, as World Wars I \& II generate two large peaks that more or less dwarf the usage frequencies reported in other decades. This observation goes hand in hand not just with the magnitude of the two World Wars, but also with the increase in the usage frequency of ``war'' in the published literature at that time. In agreement with the historical events is the estimated trend line depicted by the thick red line in panel (a). However, even after the detrending, the resulting culturomic trajectory still clearly reflects history in that the periods of World Wars I \& II stand out from the rest, as can be inferred from the curve depicted in panel (b). The Hurst parameter $H$ determined using the detrended and original data [presented in panels (c) and (d)] have similar values to each other ($H=1.09$ for the detrended data and $H=1.15$ for the original data). As a result, both classify ``war'' as either a nonstationary, on-off intermittent, or a L{\'e}vy walk-like process.

\begin{figure}
\begin{center}
\includegraphics[width=8.1cm]{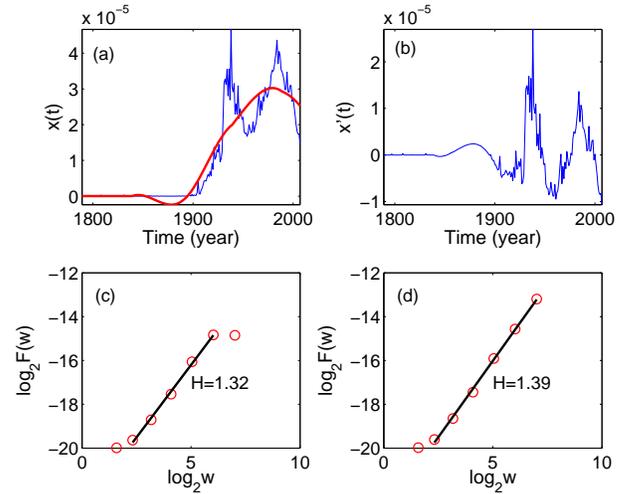}
\end{center}
\caption{Adaptive fractal analysis of the usage frequency of the 1-gram ``unemployment'' in the corpus of English books. The Hurst parameter, as obtained from the detrended data, is $H=1.32$. a) The variation of the usage frequency of ``unemployment'' with time. The blue (thin) line depicts the original data while the red (thick) line depicts the estimated trend (using a window of length $101$). b) Detrended data, \textit{i.e.} the difference between the blue and red curves. c) Best fit to the $F(w)$ versus $w$ dependence for detrended data on a double log scale yields $H=1.32$. d) Best fit to the $F(w)$ versus $w$ dependence for original data on a double log scale yields $H=1.39$.}
\label{unemployment}
\end{figure}

Another illustrative example of fractal analysis is presented in figure~\ref{unemployment}, where we examine the 1-gram ``unemployment''. A crucial distinction from ``war'', as well as all the considered natural phenomena, is that unemployment was nonexistent, or at least it was not mentioned, in the literature prior to 1900, which is clearly inferable from the original data depicted thin blue in panel (a). With the coming of age of the industrial revolution, the job market began to take shape, and with it came, rather inevitably it seems, the problem of unemployment. The trend line depicted thick red in panel (a) clearly captures this fact. Moreover, we note that the first broad peak in the plot starts at around $1930$, and thus correlates well with the Great Depression, while the second broad peak starts at around $1970$, and thus correlates with that period of US economic stagnation and high inflation that was linked with the Middle Eastern oil crisis. After detrending, the situation is of course only marginally improved (in terms of assuring a more stationary record), as can be concluded from the curve depicted in panel (b). The Hurst parameters, equalling $H=1.32$ for the detrended data (c) and $H=1.39$ for the original data (d), both clearly reflect nonstationarity, and accordingly, ``unemployment'' can be considered the result of such a process.

\begin{table}
\begin{center}
\begin{tabular}{|c|c|c|}
\multicolumn{3}{c}{}\\\hline
1-grams & \multicolumn{2}{c|} {Hurst Parameter ($H$)}  \\\cline{2-2}\cline{3-3}
                & AFA             & DFA             \\\hline
Christian       & 0.85 $\pm$ 0.05 & 0.95 $\pm$ 0.08 \\\hline
communism       & 1.32 $\pm$ 0.04 & 1.44 $\pm$ 0.05 \\\hline
crisis          & 1.15 $\pm$ 0.05 & 1.13 $\pm$ 0.08 \\\hline
democracy       & 1.18 $\pm$ 0.02 & 1.07 $\pm$ 0.07 \\\hline
education       & 1.04 $\pm$ 0.05 & 1.09 $\pm$ 0.13 \\\hline
environment     & 1.13 $\pm$ 0.04 & 1.24 $\pm$ 0.08 \\\hline
famine          & 0.74 $\pm$ 0.02 & 0.66 $\pm$ 0.06 \\\hline
malnutrition    & 1.10 $\pm$ 0.07 & 1.08 $\pm$ 0.11 \\\hline
politics        & 1.14 $\pm$ 0.03 & 0.99 $\pm$ 0.06 \\\hline
population      & 1.01 $\pm$ 0.06 & 0.98 $\pm$ 0.10 \\\hline
recession       & 1.33 $\pm$ 0.05 & 1.06 $\pm$ 0.07 \\\hline
socializing     & 1.28 $\pm$ 0.07 & 1.28 $\pm$ 0.09 \\\hline
stock           & 1.01 $\pm$ 0.05 & 0.99 $\pm$ 0.11 \\\hline
unemployment    & 1.32 $\pm$ 0.04 & 1.28 $\pm$ 0.04 \\\hline
war             & 1.09 $\pm$ 0.03 & 0.99 $\pm$ 0.12 \\\hline
\end{tabular}
\end{center}
\caption{\label{social}Hurst parameters $H$, as obtained for the detrended data of all fifteen considered 1-grams describing social phenomena. The left column lists results as obtained with the adaptive fractal analysis (AFA), while the right column lists results as obtained with the detrended fluctuation analysis (DFA). The range of values as obtained by AFA is $0.74 \leq H \leq 1.33$, with the average over all fifteen considered 1-grams equalling $\overline H=1.11$. With DFA we obtain $0.66 \leq H \leq 1.44$ and $\overline H=1.08$.}
\end{table}

As in the case with natural phenomena (see table~\ref{natural}), we also  performed the same fractal analysis as in ``war'' and ``unemployment'', along with the detrended fluctuation analysis (DFA), for thirteen other social phenomena. The results are presented in table~\ref{social}. It can be observed that the large majority of considered 1-grams have $H>1$ (AFA), which indicates that social phenomena are most likely to be either nonstationary, on-off intermittent, or L{\'e}vy walk-like process. This conclusion is obtained irrespective of whether detrending is performed or not, although the average Hurst parameter for detrended data, equalling $\overline H=1.11$ (AFA), is smaller than that obtained for original data (before detrending, not shown), which is $\overline H=1.26$. This technical discrepancy, however, is likely due to the successful removal of some level of nonstationarity that is in general characteristic of social phenomena (more so than of natural phenomena). We would like to note, however, that in general not all $H>1$ occurrences should be, by default, attributed to nonstationarity in the trajectories. While visual inspection may lend support to such a conclusion, as this was the case for results presented in figure~\ref{unemployment}, in general the $H$ value alone cannot distinguish between nonstationary, on-off intermittent or L{\'e}vy walk-like processes. In fact, the time series are too short for a robust assessment of a more precise nature of the examined social phenomena. At a glance, and since this is indeed most common, it seems convenient to attribute $H>1$ in social phenomena to nonstationarity, yet only additional future data can enable us to differentiate whether the peaks are part of an on-off intermittent process with power law distributed on and/or off events, or if they are part of a L{\'e}vy walk. Lastly, we would also like to point out that of course not \emph{all} phenomena that can be considered as social will have $H>1$. Examples include 1-grams such as ``famine'' or ``Christian'', which for the largest parts of the recorded human history were either directly related to natural phenomena (severe droughts, flooding, or other phenomena negatively affected that season's yield on vegetables, crops, grass, and animal population, hence leading to famine) or have been an integral part of the human culture for a long time (prior to the start of the culturomic trajectories). Moreover, social topics that are of little interest will not garner much attention, and are as such also unlikely to have usage frequencies with $H>1$. The social phenomena where the human factor has played a key role recently and which are reasonably popular, however, all share features that are characteristic of processes with $H>1$. In fact, it seems just to conclude that the more the social phenomena can be considered recent (unemployment, recession, democracy), the higher their Hurst parameter is likely to be (see table~\ref{social}). This agrees nicely also with the recent observation of bursts and heavy tails in human dynamics \cite{barabasi_n05}.

\section*{3. DISCUSSION}

By applying fractal analysis based on DFA and AFA to culturomic trajectories of 1-grams describing typical social and natural phenomena over the past two centuries, we have found that they obey different scaling laws. As we will discuss in what follows, our findings agree nicely with existing theory and expectations, as well as offer new interpretations as to what might be the main driving forces behind the examined phenomena.

We find that natural phenomena have properties that are typical of processes that generate persistent long-range correlations, as evidenced by the Hurst parameter being in the range $0.55 \leq H \leq 0.85$, with an average over all fifteen considered 1-grams equalling $\overline H=0.69$ (AFA). The prevalence of long-term memory in natural phenomena compels us to conjecture that the long-range correlations in the usage frequency of the corresponding terms is predominantly driven by occurrences in nature of those phenomena. Using data from five million digitized books to arrive at this understanding certainly supports the declared goal of culturomics and lends strong support to its core principles. Owing to this memory, and of course by using statistical data available, we know, based on the Gutenberg-Richter law \cite{gutenberg_54}, that in the United Kingdom, for example, an earthquake of $3.7-4.6$ on the Richter scale is likely to happen every year, an earthquake of $4.7-5.5$ is due approximately every 10 years, while an earthquake of 5.6 or larger is bound to happen every 100 years \cite{musson_03}. Similar ``statistical predictions'' are available for tsunamis and many other, if not all, natural phenomena. On a more personal level, this also agrees with how we naturally develop an understanding for the weather and related natural phenomena for the region we live in.

Social phenomena, on the other hand, have the Hurst parameter in the range $0.74 \leq H \leq 1.33$, with an average over all fifteen considered 1-grams equalling $\overline H=1.11$ (AFA). This is indicative of nonstationary processes, or stationary processes like on-off intermittency with power-law distributed on and/or off periods or L{\'e}vy walks. While our analysis does not allow distinction between these three options, it is clear that all these processes are fundamentally different from those describing natural phenomena. So while it is common to hear speculations about possible average periods regarding social phenomena -- for instance, that there may be an average period between major wars or stock market crashes -- our analysis suggests this is not the case, and that social phenomena tend to follow different scaling laws than natural phenomena. Such a difference is not unexpected, as social phenomena are, by nature, more complex than natural phenomena; the former depend on political, economic, and social forces, as well as on natural phenomena. The results of this additional complexity can be seen in our fractal analysis of a set of culturomic trajectories.

In summary, we hope to have successfully demonstrated that the data made available through the Culturomics project \cite{michel_s11}, when coupled with advanced methods of analysis, offer fascinating opportunities to explore human culture in the broadest possible sense.

\section*{APPENDIX A. METHODS}
\subsection*{A.1. Nonlinear adaptive multiscale decomposition}

Nonlinear adaptive multiscale decomposition starts by partitioning a time series into segments of length $w=2n+1$, where neighboring segments overlap by $n+1$ points, thus introducing a time scale of $\frac{w+1}{2} \tau = (n+1) \tau$, where $\tau$ is the sampling time. Each segment is then fitted with the best polynomial of order $M$. Note that $M=0$ and 1 correspond to piece-wise constant and linear fitting, respectively. We denote the fitted polynomials for the $i$-th and $(i+1)$-th segments
by $y^{(i)} (l_1)$ and $y^{(i+1)} (l_2)$, respectively, where $l_1,l_2 = 1,\cdots,2n+1$. We then define the fitting for the overlapped region as
\begin{equation}
y^{(c)} (l) = w_1 y^{(i)} (l+n) + w_2 y^{(i+1)} (l),~l=1,\cdots, n+1,
\label{trend}
\end{equation}
where $w_1 = \big (1-\frac{l-1}{n}   \big )$ and $w_2=\frac{l-1}{n}$ can be written as $(1-d_j/n)$ for $j=1,2$, and where $d_j$ denotes the distances between the point and the centers of $y^{(i)}$ and $y^{(i+1)}$, respectively. This means that the weights decrease linearly with the distance between the point and the center of the segment. Such a weighting ensures symmetry and effectively eliminates any jumps or discontinuities around the boundaries of neighboring segments. In fact, the scheme ensures that the fitting is continuous everywhere, is smooth at the non-boundary points, and has the right- and left-derivatives at the boundary. Moreover, since it can deal with an arbitrary trend without a priori knowledge, it can remove nonstationarity, including baseline drifts and motion artifacts \cite{gao_po11}, and the procedure may also be used as either high-pass or low-pass filter with superior noise-removal properties than linear filters, wavelet shrinkage, or chaos-based noise reduction schemes \cite{tung_pre11}.

\subsection*{A.2. Fractal analysis}

Based on the described adaptive decomposition, a fractal analysis can be conducted as follows. Let \{$x_1, x_2, \cdots, x_n$\} be a stationary stochastic process with mean $\overline x$ and autocorrelation function of type
\begin{equation}
r(k) \sim k^{2H-2}~~{\rm as}~~k \rightarrow \infty,
\end{equation}
where $H$ is the Hurst parameter. This is often called an increment process, and its power spectral density is $1/f^{2H-1}$. The integral of the increment process:
\begin{equation}
u(i)=\sum_{k=1}^{i} (x_k - \overline x),~~i=1,2,\cdots,n,
\label{rw}
\end{equation}
on the other hand, is called a random walk process, and its power spectral density is $1/f^{2H+1}$. Starting from an increment process, similarly to detrended fluctuation analysis \cite{peng_pre94}, we first construct a random walk process using equation~(\ref{rw}). If, however, the original data can already be classified as a random walk-like process, then this step is not necessary, although for ideal fractal processes there is no penalty even if this step is done. Next, for a window size $w$, we determine, for the random walk process $u(i)$ (or the original process if it is already a random walk-like process), a global trend $v(i),~i=1,2,\cdots, N$, where $N$ is the length of the walk. The residual, $u(i) - v(i)$, characterizes fluctuations around the global trend, and its variance yields the Hurst parameter $H$ according to
\begin{equation}
F(w) = \Big [ \frac{1}{N} \sum_{i=1}^N (u(i) - v(i))^2 \Big ]^{1/2} \sim w^{H}.
\label{ada}
\end{equation}
The validity of equation~(\ref{ada}) can be proven if one starts from an increment process with the Hurst parameter equal to $H$. Using Parseval's theorem \cite{gao_07}, the variance of the residual data corresponding to a window size $w$ may be equated to the total power $P$ in the frequency range ($f_w, f_{cutoff}$) as
\begin{equation}
P~\sim \int_{f_w}^{f_{cutoff}} \frac{1}{f^{2H+1}} {\rm d}f
~\sim~ \frac{1}{2H} \Big( w^{2H} - f_{cutoff} ^{-2H} \Big),
\label{psdp}
\end{equation}
where $f_w = 1/w$, and $f_{cutoff}$ is the highest frequency of the data. When $f_w \ll f_{cutoff}$, we see that equation~(\ref{ada}) has to be valid. In fact, the above treatment makes it clear that even if we start from a random walk process with the Hurst exponent equal to $H$, integration will give the process a spectrum of $1/f^{2H+1+2} = 1/f^{2(H+1)+1}$, and therefore, the final Hurst parameter will be simply $H+1$. This in turn indicates that there is indeed no penalty if one uses equation~(\ref{rw}) when the data are already a random walk-like process. Note that the proposed approach, if needed, can be readily extended and applied successfully to multifractal as well as higher dimensional data.

\begin{figure}
\begin{center}
\includegraphics[width=6.3cm]{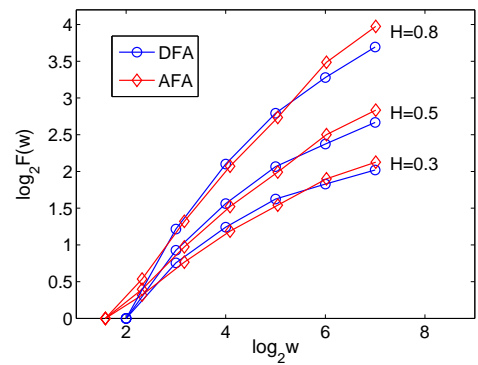}
\end{center}
\caption{Scaling analysis of fractional Gaussian noise processes of the same length as the $1$-gram data ($240$ points). Blue (circles) and red (diamonds) curves depict results as obtained by means of detrended fluctuation analysis (DFA) and adaptive fractal analysis (AFA), respectively, for three different values of $H$. It can be observed that both methods yield consistent results, regardless of the shortage of the examined time series.}
\label{method}
\end{figure}

The described fractal analysis approach, which we will refer to as adaptive fractal analysis (AFA), in general yields results that are consistent with the traditionally used detrended fluctuation analysis \cite{peng_pre94} (DFA), as can be concluded from results presented in figure~\ref{method}. Nevertheless, especially for processes having $H>1$ \cite{gao_pre06}, AFA may yield better scaling, which is why, although we analyze the culturomic trajectories with both methods, we rely on the results of AFA for final interpretation. The potential advantage of adaptive fractal analysis over detrended fluctuation analysis is due to the fact that the trend for each window of size $w$ obtained by AFA is smooth, while that obtained by DFA may change abruptly at the boundary of neighboring segments. For short nonstationary time series this may prove favorable for obtaining better scaling in the $F(w)$ versus $w$ dependence.

\begin{acknowledgments}
This research was supported in part by the US NSF grant CMMI-0825311 to Jianbo Gao, and by the Slovenian Research Agency's grant J1-4055 to Matja{\v z} Perc. The authors are grateful to Professor Johnny Lin of North Park University for many useful discussions.
\end{acknowledgments}


\begin{thebibliography}{10}

\bibitem{mantegna_00}
Mantegna, R.~N. and Stanley, H.~E.
\newblock {\em Introduction to Econophysics: Correlations \& Complexity in
  Finance}.
\newblock (Cambridge University Press, Cambridge, 2000).

\bibitem{kantz_98}
Kantz, H. and Schreiber, T.
\newblock {\em Nonlinear Time Series Analysis}.
\newblock (Cambridge University Press, Cambridge, 1998).

\bibitem{glass_88}
Glass, L. and Mackey, M.~C.
\newblock {\em From Clocks to Chaos: The Rhythms of Life}.
\newblock (Princeton University Press, Princeton, 1988).

\bibitem{goldberger_c00}
Goldberger, A.~L., Amaral, L. A.~N., Glass, L., Hausdorff, J.~M., Ivanov,
  P.~Ch., Mark, R.~G., Mietus, J.~E., Moody, G.~B., Peng, C.~K., and Stanley,
  H.~E.
\newblock Physiobank, physiotoolkit, and physionet: components of a new
  research resource for complex physiologic signals.
\newblock {\em Circulation}{ \bf 101}, 215 (2000).

\bibitem{lorenz_jas63}
Lorenz, E.~N.
\newblock Deterministic nonperiodic flow.
\newblock {\em J. Atmos. Sci.}{ \bf 20}, 130 (1963).

\bibitem{crutchfield_pr82}
Crutchfield, J.~P., Farmer, J.~D., and Huberman, B.~A.
\newblock Fluctuations and simple chaotic dynamics.
\newblock {\em Phys. Rep.}{ \bf 92}, 45 (1982).

\bibitem{eckmann_rmp85}
Eckmann, J.~P. and Ruelle, D.
\newblock Ergodic theory of strange attractors.
\newblock {\em Rev. Mod. Phys.}{ \bf 57}, 617 (1985).

\bibitem{abarbanel_96}
Abarbanel, H. D.~I.
\newblock {\em Analysis of Observed Chaotic Data}.
\newblock (Springer, New York, 1996).

\bibitem{gao_07}
Gao, J.~B., Cao, Y.~H., Tung, W.~W., and Hu, J.
\newblock {\em Multiscale Analysis of Complex Time Series: Integration of Chaos
  and Random Fractal Theory, and Beyond}.
\newblock (Wiley-Interscience, New Jersey, 2007).

\bibitem{gao_pre06}
Gao, J.~B., Hu, J., Tung, W.~W., Cao, Y.~H., Sarshar, N., and Roychowdhury,
  V.~P.
\newblock Assessment of long-range correlation in time series: How to avoid
  pitfalls.
\newblock {\em Phys. Rev. E}{ \bf 73}, 016117 (2006).

\bibitem{stratonovich_63}
Stratonovich, R.~L.
\newblock {\em Topics in the Theory of Random Noise}.
\newblock (Gordon and Breach, New York, 1963).

\bibitem{mandelbrot_82}
Mandelbrot, B.~B.
\newblock {\em The Fractal Geometry of Nature}.
\newblock (Freeman, San Francisco, 1982).

\bibitem{bunde_96}
Bunde, A. and Havlin, S.
\newblock {\em Fractals and Disordered Systems}.
\newblock (Springer, New York, 1996).

\bibitem{stanley_n88}
Stanley, H.~E. and Meakin, P.
\newblock Multifractal phenomena in physics and chemistry.
\newblock {\em Nature}{ \bf 335}, 405 (1988).

\bibitem{michel_s11}
Michel, J.~B., Shen, Y.~K., \protect{Presser Aiden}, A., Veres, A., Gray,
  M.~K., \protect{The Google Books Team}, Pickett, J.~P., Hoiberg, D., Clancy,
  D., Norvig, P., Orwant, J., Pinker, S., Nowak, M.~A., and \protect{Lieberman
  Aiden}, E.
\newblock Quantitative analysis of culture using millions of digitized books.
\newblock {\em Science}{ \bf 331}, 176 (2011).

\bibitem{gonzales_n08}
Gonz{\'a}lez, M.~C., Hidalgo, C.~A., and Barab{\'a}si, A.~L.
\newblock Understanding individual human mobility patterns.
\newblock {\em Nature}{ \bf 453}, 779 (2008).

\bibitem{song_np10}
Song, C., Koren, T., Wang, P., and Barab{\'a}si, A.~L.
\newblock Modelling the scaling properties of human mobility.
\newblock {\em Nature Physics}{ \bf 6}, 818 (2010).

\bibitem{song_s10}
Song, C., Qu, Z., Blumm, N., and Barab{\'a}si, A.~L.
\newblock Limits of predictability in human mobility.
\newblock {\em Science}{ \bf 327}, 1018 (2010).

\bibitem{balcan_pnas09}
Balcan, D., Colizza, V., Gon{\c c}alves, B., Hu, H., Ramasco, J.~J., and
  Vespignani, A.
\newblock Multiscale mobility networks and the spatial spreading of infectious
  diseases.
\newblock {\em Proc. Natl. Acad. Sci. USA}{ \bf 106}, 21484 (2009).

\bibitem{meloni_pnas09}
Meloni, S., Arenas, A., and Moreno, Y.
\newblock Traffic-driven epidemic spreading in finite-size scale-free networks.
\newblock {\em Proc. Natl. Acad. Sci. USA}{ \bf 106}, 16897 (2009).

\bibitem{sanz_pre10}
Sanz, J., Flor{\'i}a, L.~M., and Moreno, Y.
\newblock Spreading of persistent infections in heterogeneous populations.
\newblock {\em Phys. Rev. E}{ \bf 81}, 056108 (2010).

\bibitem{meloni_sr11}
Meloni, S., Perra, N., Arenas, A., G{\'o}mez, S., Moreno, Y., and Vespignani,
  A.
\newblock Modeling human mobility responses to the large-scale spreading of
  infectious diseases.
\newblock {\em Scientific Reports}{ \bf 1}, 62 (2011).

\bibitem{hu_pnas09}
Hu, H., Myers, S., Colizza, V., and Vespignani, A.
\newblock Wifi networks and malware epidemiology.
\newblock {\em Proc. Natl. Acad. Sci. USA}{ \bf 106}, 1318 (2009).

\bibitem{wang_s09}
Wang, P., Gonz{\'a}lez, M., Hidalgo, C.~A., and Barab{\'a}si, A.~L.
\newblock Understanding the spreading patterns of mobile phone viruses.
\newblock {\em Science}{ \bf 324}, 1071 (2009).

\bibitem{ratkiewicz_prl10}
Ratkiewicz, J., Fortunato, S., Flammini, A., Menczer, F., and Vespignani, A.
\newblock Characterizing and modeling the dynamics of online popularity.
\newblock {\em Phys. Rev. Lett.}{ \bf 105}, 158701 (2010).

\bibitem{holt_pone11}
Borge-Holthoefer, J., Rivero, A., Garcia, I., Cauhe, E., Ferrer, A., Ferrer,
  D., Francos, D., Iniguez, D., Perez, M.~P., Ruiz, G., Sanz, F., Serrano, F.,
  Vinas, C., Tarancon, A., and Moreno, Y.
\newblock Structural and dynamical patterns on online social networks: The
  spanish may 15th movement as a case study.
\newblock {\em PLoS ONE}{ \bf 6}, e23883 (2011).

\bibitem{lieberman_n07}
Lieberman, E., Michel, J.~B., Jackson, J., Tang, T., and Nowak, M.~A.
\newblock Quantifying the evolutionary dynamics of language.
\newblock {\em Nature}{ \bf 449}, 713 (2007).

\bibitem{puglisi_pnas08}
Puglisi, A., Baronchelli, A., and Loreto, V.
\newblock Cultural route to the emergence of linguistic categories.
\newblock {\em Proc. Natl. Acad. Sci. USA}{ \bf 105}, 7936 (2008).

\bibitem{loreto_jsm11}
Loreto, V., Baronchelli, A., Mukherjee, A., Puglisi, A., and Tria, F.
\newblock Statistical physics of language dynamics.
\newblock {\em J. Stat. Mech.} P04006 (2011).

\bibitem{radicchi_po11}
Radicchi, F.
\newblock Who is the best player ever? A complex network analysis of the
  history of professional tennis.
\newblock {\em PLoS ONE}{ \bf 6}, e17249 (2011).

\bibitem{evans_s11}
Evans, J.~A. and Foster, J.~G.
\newblock Metaknowledge.
\newblock {\em Science}{ \bf 331}, 721 (2011).

\bibitem{lazer_s09}
Lazer, D., Pentland, A., Adamic, L.~A., Aral, S., Barab{\'a}si, A.~L., Brewer,
  D., Christakis, N., Contractor, N., Fowler, J., Gutmann, M., Jebara, T.,
  King, G., Macy, M., Roy, D., and \protect{Van Alstyne}, M.
\newblock Computational social science.
\newblock {\em Science}{ \bf 323}, 721 (2009).

\bibitem{stanley_87}
Stanley, H.~E.
\newblock {\em Introduction to Phase Transitions and Critical Phenomena}.
\newblock (Oxford University Press, Oxford, 1987).

\bibitem{press_cmpc78}
Press, W.~H.
\newblock Flicker noises in astronomy and elsewhere.
\newblock {\em Comments Astrophys.}{ \bf 7}, 103 (1978).

\bibitem{bak_prl87}
Bak, P., Tang, C., and Wiesenfeld, K.
\newblock Self-organized criticality: An explanation of \protect{$1/f$} noise.
\newblock {\em Phys. Rev. Lett.}{ \bf 59}, 381 (1987).

\bibitem{bak_96}
Bak, P.
\newblock {\em How Nature Works: The Science of Self-Organized Criticality}.
\newblock (Copernicus, New York, 1996).

\bibitem{gao_pla03}
Gao, J.~B., Cao, Y.~H., and Lee, J.~M.
\newblock Principal component analysis of \protect{$1/f$} noise.
\newblock {\em Phys. Lett. A}{ \bf 314}, 392 (2003).

\bibitem{voss_prl92}
Voss, R.~F.
\newblock Evolution of long-range fractal correlations and 1/f noise in dna
  base sequences.
\newblock {\em Phys. Rev. Lett.}{ \bf 68}, 3805 (1992).

\bibitem{peng_n92}
Peng, C.~K., Buldyrev, S.~V., Goldberger, A.~L., Havlin, S., Sciortino, F.,
  Simons, M., and Stanley, H.~E.
\newblock Long--range correlations in nucleotide sequences.
\newblock {\em Nature}{ \bf 356}, 168 (1992).

\bibitem{gilden_s95}
Gilden, D.~L., Thornton, T., and Mallon, M.~W.
\newblock 1/f noise in human cognition.
\newblock {\em Science}{ \bf 267}, 1837 (1995).

\bibitem{chen_prl97}
Chen, Y., Ding, M., and \protect{Scott Kelso}, J.~A.
\newblock {\em Phys. Rev. Lett.}{ \bf 79}, 4501 (1997).

\bibitem{collins_prl94}
Collins, J.~J. and \protect{De Luca}, C.~J.
\newblock Random walking during quiet standing.
\newblock {\em Phys. Rev. Lett.}{ \bf 73}, 764 (1994).

\bibitem{ivanov_n96}
Ivanov, P.~Ch., Rosenblum, M.~G., Peng, C.~K., Mietus, J., Havlin, S., Stanley,
  H.~E., and Goldberger, A.~L.
\newblock Scaling behaviour of heartbeat intervals obtained by wavelet-based
  time-series analysis.
\newblock {\em Nature}{ \bf 383}, 323 (1996).

\bibitem{amaral_prl98}
Amaral, L. A.~N., Goldberger, A.~L., Ivanov, P.~Ch., and Stanley, H.~E.
\newblock Scale-independent measures and pathologic cardiac dynamics.
\newblock {\em Phys. Rev. Lett.}{ \bf 81}, 2388 (1998).

\bibitem{ivanov_n99}
Ivanov, P.~Ch., Rosenblum, M.~G., Amaral, L. A.~N., Struzik, Z.~R., Havlin, S.,
  Goldberger, A.~L., and Stanley, H.~E.
\newblock Multifractality in human heartbeat dynamics.
\newblock {\em Nature}{ \bf 399}, 461 (1999).

\bibitem{galvan_prl01}
\protect{Bernaola-Galvan}, P., Ivanov, P.~Ch., Amaral, L. A.~N., and Stanley,
  H.~E.
\newblock Scale invariance in the nonstationarity of human heart rate.
\newblock {\em Phys. Rev. Lett.}{ \bf 87}, 168105 (2001).

\bibitem{wolf_pa97}
Wolf, M.
\newblock 1/f noise in the distribution of prime numbers.
\newblock {\em Physica A}{ \bf 241}, 493 (1997).

\bibitem{peng_pre94}
Peng, C.~K., Buldyrev, S.~V., Havlin, S., Simons, M., and Stanley, H.~E.
\newblock Mosaic organization of DNA nucleotides.
\newblock {\em Phys. Rev. E}{ \bf 49}, 1685 (1994).

\bibitem{hu_pre01}
Hu, K., Ivanov, P.~Ch., Chen, Z., Carpena, P., and Stanley, H.~E.
\newblock Effect of trends on detrended fluctuation analysis.
\newblock {\em Phys. Rev. E}{ \bf 64}, 011114 (2001).

\bibitem{stanley_pa02}
Stanley, H.~E., Kantelhardt, J.~W., Zschiegner, S.~A.,
  \protect{Koscielny-Bunde}, E., Havlin, S., and Bunde, A.
\newblock Multifractal detrended fluctuation analysis of nonstationary time
  series.
\newblock {\em Physica A}{ \bf 316}, 87 (2002).

\bibitem{chen_pre05}
Chen, Z., Hu, K., Carpena, P., \protect{Bernaola-Galvan}, P., Stanley, H.~E.,
  and Ivanov, P.~Ch.
\newblock Effect of nonlinear filters on detrended fluctuation analysis.
\newblock {\em Phys. Rev. E}{ \bf 71}, 011104 (2005).

\bibitem{hu_jsm09}
Hu, J., Gao, J., and Wang, X.
\newblock Multifractal analysis of sunspot time series: the effects of the
  11-year cycle and fourier truncation.
\newblock {\em J. Stat. Mech.} P02066 (2009).

\bibitem{barabasi_n05}
Barab{\'a}si, A.~L.
\newblock The origin of bursts and heavy tails in humans dynamics.
\newblock {\em Nature}{ \bf 435}, 207 (2005).

\bibitem{gutenberg_54}
Gutenberg, B. and Richter, C.
\newblock {\em Seismicity of the Earth and Associated Phenomena}.
\newblock (Princeton University Press, Princeton, 1954).

\bibitem{musson_03}
Musson, R.
\newblock Seismicity and earthquake hazard in the UK.
\newblock {\em http://www.quakes.bgs.ac.uk/} (2003).

\bibitem{gao_po11}
Gao, J.~B., Hu, J., and Tung, W.~W.
\newblock Facilitating joint chaos and fractal analysis of biosignals through
  nonlinear adaptive filtering.
\newblock {\em PLoS ONE}{ \bf 6}, e24331 (2011).

\bibitem{tung_pre11}
Tung, W.~W., Gao, J., Hu, J., and Yang, L.
\newblock Recovering chaotic signals in heavy noise environments.
\newblock {\em Phys. Rev. E}{ \bf 83}, 046210 (2011).

\end{thebibliography}
\end{document}